\documentclass[12pt,a4paper]{article}
\usepackage{a4wide}
\usepackage{latexsym}
\usepackage{epsf}
\usepackage{amssymb}

\begin{document}

\begin{flushright}
\small
IFT-UAM/CSIC-01-07\\
KUL-TF-2001/9\\
{\bf hep-th/0103244}\\
March 28th 2001
\normalsize
\end{flushright}

\begin{center}
%title
\vspace{.7cm}

{\LARGE {\bf Type~0 T-Duality and the Tachyon Coupling}}\\
%\vspace{.3cm}
%{\LARGE {\bf Coupling}}
\vspace{1.3cm}
%authors
%
{\bf\large Patrick Meessen}${}^{\diamondsuit}$
\footnote{E-mail: {\tt Patrick.Meessen@fys.kuleuven.ac.be}}
{\bf\large and Tom\'as Ort\'{\i}n}${}^{\spadesuit\clubsuit}$
\footnote{E-mail: {\tt tomas@leonidas.imaff.csic.es}}
\vskip 1truecm

${}^{\diamondsuit}$\ {\it Instituut voor Theoretische
Fysica, Katholieke Universiteit Leuven, Celestijnenlaan~200D, B-3001
Leuven, Belgium.}

\vskip 0.2cm 
${}^{\spadesuit}$\ {\it Instituto de F\'{\i}sica Te\'orica, C-XVI,
Universidad Aut\'onoma de Madrid \\
E-28049-Madrid, Spain}

\vskip 0.2cm
${}^{\clubsuit}$\ {\it I.M.A.F.F., C.S.I.C., Calle de Serrano 113 bis\\ 
E-28006-Madrid, Spain}
\vspace{.7cm}
%%%%%%%%%%%%%%%%%%%%%%%%%%%%%%%%%%%%%%%%%%%%%%%%%%%%%%%%%%%%%%%%%%%%%%

{\bf Abstract}
\end{center}
\begin{quotation}
  \small We consider the T-duality relations between Type~0A and 0B
  theories, and show that this constraints the possible couplings of
  the tachyon to the RR-fields. Due to the `doubling' of the RR sector
  in Type~0 theories, we are able to introduce a democratic
  formulation for the Type~0 effective actions, in which there is no
  Chern-Simons term in the effective action.  Finally we discuss how
  to embed Type~II solutions into Type~0 theories.
\end{quotation}
\newpage
\pagestyle{plain}
%%%%%%%%%%%%%%%%%%%%%%%%%%%%%%%%%%%%%%%%%%%%%%%%%%%%%%%%%%%%%%%%%%%%%%
%%%%%%%%%%%%%%%%%%%%%%%%%%%%%%%%%%%%%%%%%%%%%%%%%%%%%%%%%%%%%%%%%%%%%%
%%%%%%%%%%%%%%%%%%%%%%%%%%%%%%%%%%%%%%%%%%%%%%%%%%%%%%%%%%%%%%%%%%%%%%
%%%%%%%%%%%%%%%%%%%%%%%%%%%%%%%%%%%%%%%%%%%%%%%%%%%%%%%%%%%%%%%%%%%%%%
\section*{Introduction}
%%%%%%%%%%%%%%%%%%%%%%%%%%%%%%%%%%%%%%%%%%%%%%%%%%%%%%%%%%%%%%%%%%%%%%
Most 10-dimensional non-supersymmetric superstring theories are
plagued with tachyons, possibly endangering the consistency of the
theory.  This not-withstanding, they have become an active field of
research mainly due to lessons learned from duality relations in the
supersymmetric theories. One of the most studied examples, and subject
of this paper, are the so-called Type~0 theories.
\par
Type~0 theories can be obtained by a diagonal GSO projection on the
superstring spectrum or by orbifolding the corresponding Type~II
theories by $(-)^{F_{s}}$, the total target space fermion number
\cite{kn:DiHa}.  Note that there are two Type~IIB theories, denoted
IIB$_{+}$ and IIB$_{-}$,\footnote{These are denoted IIB and
  IIB$^{\prime}$ {\em resp.}  in \cite{joepboek}.} which are related
by spacetime parity and lead to the same Type~0B theory.  {}From the
supergravity point of view, IIB$_{+}$ differs from IIB$_{-}$ in that
IIB$_{+}$ has a selfdual 5-form field strength whereas IIB$_{-}$ has an
anti-selfdual 5-form field strength.  Similar statements can be made
for the Type~IIA$_{\pm}$ theories and its relation to the unique
Type~0A theory.
\par
In the notation of \cite{joepboek}, the spectrum of the Type~0
theories are represented as
\begin{eqnarray}
0B &:& 
(NS_{-},NS_{-})\oplus (NS_{+},NS_{+})\oplus 
(R_{+},R_{+})\oplus (R_{-},R_{-}) \; , \nonumber \\
0A &:& 
(NS_{-},NS_{-})\oplus (NS_{+},NS_{+})\oplus 
(R_{+},R_{-})\oplus (R_{-},R_{+}) \; ,
\end{eqnarray}
which then consists of a tachyon, the string common sector and a
doubling, w.r.t. the analogous Type~II theory, of the RR-fields.
Since there is a doubling of the RR-fields, there is also a doubling
of the D-brane content in the Type~0 theories.

In Ref.~\cite{kn:KlTs} the lowest order field contributions to the
Type~0B string effective action were calculated from the appropriate
string scattering amplitudes and, in our conventions \cite{kn:MO}, and
using the diagonal basis of RR fields
$\{\hat{C}_{(p+1)}^{+},\hat{C}_{(4)}^{+},
\hat{C}_{(p+1)}^{-}\}_{p=-1,1}$, the action reads

\begin{equation}
\label{eq:0Baction-1}
\begin{array}{rcl}
\hat{S} & = & {\displaystyle\int} d^{10}\hat{x}\sqrt{|\hat{g}|}\,
\left\{e^{-2\hat{\varphi}} \left[ \hat{R} -4(\partial\hat{\varphi})^{2}
+\frac{1}{2\cdot 3!}\hat{\cal H}^{2}+\frac{1}{2}(\partial\hat{\cal T})^{2}
-V(\hat{\cal T})\right] \right.\\
& & \\
& & 
\hspace{1cm}
+f_{+}(\hat{\cal T})
\left[\frac{1}{2}\left(\hat{G}_{(1)}^{+}\right)^{2} 
+\frac{1}{2\cdot 3!}\left(\hat{G}_{(3)}^{+}\right)^{2}
+\frac{1}{2\cdot 5!}\left(\hat{G}_{(5)}^{+}\right)^{2}
\right]\\
& & \\
& & 
\hspace{1cm}
\left.
+f_{-}(\hat{\cal T})
\left[\frac{1}{2}\left(\hat{G}_{(1)}^{-}\right)^{2} 
+\frac{1}{2\cdot 3!}\left(\hat{G}_{(3)}^{-}\right)^{2}
\right]\right\}\, ,\\
\end{array}
\end{equation}

\noindent where the field strengths are defined by

\begin{equation}
  \label{eq:10dfieldstrengths}
  \hat{\mathcal{H}}\;=\; d\hat{\mathcal{B}}\hspace{.3cm},\hspace{.3cm}
  \hat{G}_{(n)}^{\pm}\;=\; d\hat{C}_{(n-1)}^{\pm} \; ,
\end{equation}

\noindent the tachyon potential is

\begin{equation}
V(\hat{\cal T})
={\textstyle\frac{1}{2}}m^{2}\hat{\cal T}^{2}-4c_{1}\hat{\cal T}^{4}\, ,
\hspace{1cm}
m^{2} =-2/\ell_{s}^{2}\, ,
\end{equation}

\noindent where $\ell_{s}=\sqrt{\alpha^{\prime}}$ 
is the string length.  It was later argued in Ref.~\cite{kn:Ts2} that
there should be no such a potential in Type~0 superstring effective
actions. However, since the tachyon is inert under T-duality, the
value of the tachyon potential $V$ will be immaterial in most of our
discussion.

Finally, the functions $f_{\pm}(\hat{\cal T})$ are given by

\begin{equation}
f_{\pm}(\hat{\cal T})=1\pm\sqrt{2}\hat{\cal T}+\hat{\cal T}^{2}+
\mathcal{O}\left(\hat{\mathcal{T}}^{4}\right) \, .
\end{equation}

The RR fields are combinations of the $(R_{+},R_{+})$ and $(R_{-},R_{-})$
fields, denoted by $C$ {\em resp.} $\overline{C}$,
that diagonalize the kinetic terms:

\begin{equation}
\sqrt{2}\ \hat{C}_{(2n)}^{\pm} = \hat{C}_{(2n)}\pm 
                                 \hat{\overline{C}}_{(2n)}\, ,
\end{equation}

\noindent and we will denote a brane charged w.r.t. $C^{+}_{(p+1)}$
($C^{-}_{(p+1)}$) by a $Dp_{+}$-brane ($Dp_{-}$-brane {\em resp.}). 

The fields $\hat{C}_{(4)}$ and $\hat{\overline{C}}_{(4)}$ have self- 
and anti-selfdual field strengths and deserve further discussion. In
principle, as in the Type~IIB$_{\pm}$ cases, it is not possible to
write a kinetic term for neither of them separately without the help
of auxiliary fields or without breaking covariance. Combining
them, however, it is possible to write a kinetic term of the form
$\hat{G}_{(5)}\hat{\overline{G}}_{(5)}$. From this term one recovers a
standard-looking equation of motion, but not self- or
anti-self-duality which still has to be imposed by hand. It is easy to
convince oneself that this cannot be done consistently in the presence
of coupling to the tachyon: the equation of motion

\begin{equation}
d\left( f_{+} {}^{\star}\hat{G}_{(5)}\right) \;=\; 0\, ,  
\end{equation}

\noindent should give the Bianchi identity under the duality 
transformation, which has to be

\begin{equation}
f_{+} {}^{\star}\hat{G}_{(5)} = \hat{G}_{(5)}\, ,
\end{equation}

\noindent which is clearly inconsistent for non-constant $f_{+}$.

Another possibility is to combine both of them into a completely
unconstrained 5-form field strength $\hat{G}^{+}_{(5)}$ with standard
kinetic term, as it has been done here in the action
Eq.~(\ref{eq:0Baction-1}). All we require is that it leads to the
right equations of motion associated to the propagators that can be
calculated from string amplitudes. Defining

\begin{equation}
\label{eq:constraint}
 \hat{G}^{-}_{(5)}\equiv f_{+} {}^{\star}\hat{G}^{+}_{(5)}\, .
\end{equation}

\noindent we can immediately find an alternative to the action
Eq.~(\ref{eq:0Baction-1}) in which all the kinetic terms of the fields
with a minus superscript have a factor $f_{-}\left(\hat{\cal
    T}\right)$ except for $\hat{G}^{-}_{(5)}$ that carries a factor
$f_{+}^{-1}\left(\hat{\cal T}\right)$.

Yet another possibility, which we will use later on, is to write an
almost standard kinetic terms for $\hat{C}^{+}_{(4)}$ and
$\hat{C}^{-}_{(4)}$ with the understanding that self- and
anti-self-duality have to be imposed on the subsequent equations of
motion. In this case, the kinetic term would be

\begin{equation}
\label{eq:NSDkineticterms}
\int d^{10}\hat{x}\, \sqrt{|\hat{\jmath}|}\, 
\left\{f_{+}{\textstyle\frac{1}{4\cdot 5!}}
\left(\hat{G}^{+}_{(5)}\right)^{2}
+f_{+}^{-1}{\textstyle\frac{1}{4\cdot 5!}}
\left(\hat{G}^{-}_{(5)}\right)^{2}\right\}\, ,
\end{equation}

\noindent and we would impose Eq.~(\ref{eq:constraint}) as a constraint.
This {\em non-selfdual} (NSD) action would be a generalization of the
Type~IIB one \cite{kn:BBO,kn:MO}. Eliminating the $\hat{G}^{-}_{(5)}$
combination with the above constraint would take us to
Eq.~(\ref{eq:0Baction-1}). Eliminating $\hat{G}^{+}_{(5)}$ would give
us the  alternative action in terms of $\hat{G}^{-}_{(5)}$.

%%%%%%%%%%%%%%%%%%%%%%%%%%%%%%%%%%%%%%%%%%%%%%%%%%%%%%%%%%%%%%%%%%%%%%%

A further remark must be made: the Type~0 theories have to be
invariant under a $\mathbb{Z}_{2}$ group associated to $(-)^{f_{L}}$,
the worldsheet fermion number.  This implies that the effective action
should be invariant under the transformation
$\hat{\mathcal{T}}\rightarrow -\hat{\mathcal{T}}$ combined with an
interchange of the $^{+}$- and $^{-}$ fields.  A quick look at
Eq.~(\ref{eq:0Baction-1}) then reveals that this can only be true if
\begin{equation}
  \label{eq:relatie1}
  f_{+}\left(\hat{\mathcal{T}}\right) \;=\;
        f_{-}\left( -\hat{\mathcal{T}}\right) \; .
\end{equation}
Furthermore, since in the interchange of the $^{+}$- and $^{-}$ fields
$\hat{G}^{+}_{(5)}$ is transformed into $\hat{G}^{-}_{(5)}$, it is
clear that the action is not strictly invariant under this
transformation. Actually, it takes us to the alternative
action\footnote{We could also say that the transformation has to be
  supplemented by a dualization of $\hat{G}^{-}_{(5)}$ to be a
  symmetry of the action.} but with the 5-form kinetic term carrying a
tachyon factor $f_{+}\left(-\hat{\mathcal{T}}\right)$ instead of 
$f_{+}^{-1}\left(\hat{\mathcal{T}}\right)$. This implies that 

\begin{equation}
f_{+}\left(-\hat{\mathcal{T}}\right) \;=\;
        f_{+}^{-1}\left(\hat{\mathcal{T}}\right) 
\;=\; f_{-}\left(\hat{\mathcal{T}}\right)\; .
\end{equation}

These constraints, which will also coincide with the constraints
coming from T-duality, determine to some extent the form of the
functions $f$, as we will discuss later.

%%%%%%%%%%%%%%%%%%%%%%%%%%%%%%%%%%%%%%%%%%%%%%%%%%%%%%%%%%%%%%%%%%%%%%%

Due to the similarity of Type~0B with Type~IIB we expect more terms in
the RR field strengths and a Chern-Simons term in the action. These
terms could in principle be determined from more complicated string
amplitudes, but we are going to try to determine as many as we can of
these additional terms by imposing T-duality between the Type~0B and
Type~0A string effective actions using dimensional reduction as in
Refs.~\cite{kn:BHO,kn:BRGPT,kn:MO}. This is our main goal.  The
Type~0A string effective action has not been calculated from first
principles as yet. However, it is clear that the tachyon-independent
part of the NSNS sector effective action (the so-called ``common
sector'') is identical to the Type~0B one.  Furthermore, it is also
clear that T-duality acts on this sector according to the usual
Buscher rules\footnote{In closed string theory, T-duality is always a
  symmetry that interchanges momentum and winding modes associated to
  a given compact direction whose radius is simultaneously inverted.
  It is worth remarking that the string effective action does not
  contain any field associated to these modes (they are massive). One
  could take into account massive Kaluza-Klein (momentum) modes
  arising in the compactification of the massless fields contained in
  the effective action, but there is no known way to take into account
  winding modes, which are stringy (not field-theoretical) objects.
  Given this fact, one may wonder how, if at all, can the string
  effective action give a description of T-duality. The answer lies in
  the observation that all Kaluza-Klein modes are charged with respect
  to the massless Kaluza-Klein vector coming from the metric while all
  winding modes are charged with respect to the {\it winding} vector
  coming from the Kalb-Ramond 2-form. The interchange of momentum and
  winding modes implies the interchange of the Kaluza-Klein and
  winding vectors of the string effective action. Furthermore, the
  inversion of the radius is expressed in the effective action as the
  inversion of the Kaluza-Klein scalar that measures that radius.  The
  transformation of the remaining massless fields follows from these
  and from covariance. This is the content of the Buscher rules and
  this is why they have to take the same form in the NSNS sector of
  any closed string theory effective action.}  \cite{kn:Bu} which also
implies that the tachyon is invariant under T-duality. We are going to
show that these facts, plus our knowledge of the field content of the
Type~0A theory and its T-duality relation to the Type~0B theory is
enough to determine the effective action of both the Type~0A and
Type~0B theories to identical orders in the fields.

In the next section we are going to reduce first the Type~0B
action Eq.~(\ref{eq:0Baction-1}) to nine dimensions. From the form of
the 9-dimensional action plus the T-duality invariance of the tachyon
field we will immediately be able to derive an effective action for
the Type~0A theory, including tachyon couplings about which we will
obtain more information. Next, we will notice that we need additional
terms in the RR field strengths to establish T-duality with the
Type~0A effective action. This will follow from our knowledge of
Buscher's rules in the NSNS sector.  The introduction of the new terms
on the RR field strengths will also force us to introduce a
Chern-Simons term.

%%%%%%%%%%%%%%%%%%%%%%%%%%%%%%%%%%%%%%%%%%%%%%%%%%%%%%%%%%%%%%%%%%%%%%
\section{The Type~0A action and T-duality}
\label{sec:type0-9d}
%%%%%%%%%%%%%%%%%%%%%%%%%%%%%%%%%%%%%%%%%%%%%%%%%%%%%%%%%%%%%%%%%%%%%%
As was said above, the Type~0A action has not been calculated from
first principles, although such an action was proposed in
Ref.~\cite{kn:FKM}. In this section we will use T-duality as a
guideline for the construction of the Type~0A effective action.  We
will leave the construction of the massive theory for section
(\ref{sec:DemAction}).
\subsection{Reduction to $d=9$ of the Type~0B Effective Action}
Our Kaluza-Klein Ansatz to reduce the the Type~0B action
Eq.~(\ref{eq:0Baction-1}) in the direction of the coordinate $y=x^{9}$
will be similar to the one used in establishing Type~IIA/B T-duality
in Ref.~\cite{kn:MO} (identical in the NSNS sector, actually). The
relation between the 10-dimensional fields

\begin{equation}
\{\hat{\jmath},
\hat{\cal B},\hat{\varphi},\hat{\cal T}, 
\hat{C}^{+}_{(0)},\hat{C}^{-}_{(0)},\hat{C}^{+}_{(2)},
\hat{C}^{-}_{(2)},\hat{C}^{+}_{(4)}\}\, ,
\end{equation}

\noindent and the 9-dimensional fields 

\begin{equation}
\{g,B,A^{(1)},A^{(2)},k,\phi,T,
C^{+}_{(0)},C^{+}_{(1)},C^{+}_{(2)},C^{+}_{(3)},C^{+}_{(4)},
C^{-}_{(0)},C^{-}_{(1)},C^{-}_{(2)},\}\, ,  
\end{equation}

\noindent is, in the NSNS sector

\begin{equation}
\begin{array}{rclrcl}
\hat{\jmath}_{\mu\nu} & = & 
g_{\mu\nu}-k^{-2}A^{(2)}{}_{\mu}A^{(2)}{}_{\nu}\, ,
\hspace{1cm}&
\hat{\cal B}_{\mu\nu} & = & 
B_{\mu\nu}+A^{(1)}{}_{[\mu}A^{(2)}{}_{\nu]}\, ,
\\
& & & & &
\\
\hat{\jmath}_{\mu\underline{y}} & = & -k^{-2}A^{(2)}{}_{\mu}\, , &
\hat{\cal B}_{\mu\underline{y}} & = & A^{(1)}{}_{\mu}\, ,
\\
& & & & &
\\
\hat{\jmath}_{\underline{yy}} & = & -k^{-2}\, , &
\hat{\varphi}               & = & \phi -\frac{1}{2}\log{k}\, ,
\\
& & & & & 
\\
\hat{\cal T} & = & T\, ,& & & \\
\end{array}
\label{eq:9dNSNS0B}
\end{equation}

\noindent and, in the RR sector

\begin{equation}
\begin{array}{lcl}
\hat{C}^{\pm}_{(2n)\, \mu_{1}\cdots\mu_{2n}} & = & 
C^{\pm}_{(2n)\, \mu_{1}\cdots\mu_{2n}}
-2nA^{(2)}{}_{[\mu_{1}} C^{\pm}_{(2n-1)\, \mu_{2}\cdots\mu_{2n}]}\, ,\\
& & \\
\hat{C}^{\pm}_{(2n)\, \mu_{1}\cdots\mu_{2n-1}\underline{y}} & = & 
-C^{\pm}_{(2n-1)\, \mu_{1}\cdots\mu_{2n-1}}\, .\\ 
\end{array}
\label{eq:9dRR0B-1}
\end{equation}

The field strengths are related, in flat indices, by

\begin{equation}
\left\{
\begin{array}{rcl}
\hat{\cal H}_{abc} & = & H_{abc}\, ,\\
& & \\
\hat{\cal H}_{aby} & = & k F^{(1)}{}_{ab}\, ,\\
\end{array}
\right.
\end{equation}

\noindent where

\begin{equation}
\label{eq:d9NSNSfieldstrengths}
\left\{
\begin{array}{rcl}
H & = & dB  -\frac{1}{2}A^{(1)}F^{(2)}
-\frac{1}{2}A^{(2)}F^{(1)}\, ,\\
& & \\
F^{(1,2)} & = & d A^{(1,2)}\, ,\\
\end{array}
\right.
\end{equation}

\noindent in the NSNS sector and by 

\begin{equation}
\left\{
\begin{array}{rcl}
\hat{G}^{\pm}_{(2n+1)\, a_{1}\cdots a_{2n+1}} & = & 
G^{\pm}_{(2n+1)\, a_{1}\cdots a_{2n+1}}\, ,\\
& & \\
\hat{G}^{\pm}_{(2n+1)\, a_{1}\cdots a_{2n}y} & = & 
-kG^{\pm}_{(2n)\, a_{1}\cdots a_{2n}}\, ,
\end{array}
\right.
\end{equation}

\noindent where

\begin{equation}
\label{eq:9dRRfieldstrengths0B-1}
\left\{
\begin{array}{lcl}
G^{\pm}_{(2n+1)} & = & dC^{\pm}_{(2n)} +F^{(2)} C^{\pm}_{(2n-1)}\, ,\\
& & \\
G^{\pm}_{(2n)} & = & dC^{\pm}_{(2n-1)}\, ,\\
\end{array}
\right.
\end{equation}

\noindent in the RR sector. The reduced action is 

\begin{equation}
\begin{array}{rcl}
S & = & {\displaystyle\int} d^{9}x\,  \sqrt{|g|}\ 
\left\{ e^{-2\phi}
\left[R -4(\partial\phi)^{2} +{\textstyle\frac{1}{2\cdot 3!}}H^{2}
+(\partial\log{k})^{2} 
\right.
\right.\\
& & \\
& &
\hspace{3.5cm}
\left.
-{\textstyle\frac{1}{4}}k^{2} \left(F^{(1)}\right)^{2} 
-{\textstyle\frac{1}{4}}k^{-2}\left(F^{(2)}\right)^{2}
+\textstyle{\frac{1}{2}}\left(\partial T\right)^{2} -V(T)
\right]
\\
& & \\
& & 
\hspace{2cm}
+f_{+}(T)
\left[\frac{1}{2}k^{-1}\left(G_{(1)}^{+}\right)^{2} 
-\frac{1}{4}k\left(G_{(2)}^{+}\right)^{2} 
+\frac{1}{2\cdot 3!}k^{-1}\left(G_{(3)}^{+}\right)^{2}
\right.
\\
& & \\
& & 
\hspace{3cm}
\left.
-\frac{1}{2\cdot 4!}k\left(G_{(4)}^{+}\right)^{2}
+\frac{1}{2\cdot 5!}k^{-1}\left(G_{(5)}^{+}\right)^{2}
\right]\\
& & \\
& & 
\hspace{2cm}
\left.
+f_{-}(T)
\left[
\frac{1}{2}k^{-1}\left(G_{(1)}^{-}\right)^{2} 
-\frac{1}{4}k\left(G_{(2)}^{-}\right)^{2} 
+\frac{1}{2\cdot 3!}k^{-1}\left(G_{(3)}^{-}\right)^{2}
\right]
\right\}\, .\\
\end{array}
\label{eq:0Baction9d-1}
\end{equation}

%%%%%%%%%%%%%%%%%%%%%%%%%%%%%%%%%%%%%%%%%%%%%%%%%%%%%%%%%%%%%%%%%%%%%%

\subsection{The Type~0A Effective Action and its Reduction to $d=9$}

We should compare the above action with the dimensionally reduced
Type~0A effective action which we do not know in detail. Let us
summarize our knowledge of this action: first of all, it contains the
same 10-dimensional NSNS fields as the Type~0B action and all of them
(except, possibly, the tachyon) appear in it in identical fashion.
This implies that the T-duality rules in this sector will be Buscher's
and also implies that the tachyon will appear also in the same form
and will be invariant under T-duality (its reduction is trivial).

As for the RR fields, the Type~0A string effective action contains 2
1-forms, and 2 3-forms:\footnote{There must be a massive extension of
  the Type~0A theory, with two constant
  field strengths $\hat{G}_{(0)}$ and $\hat{\overline{G}}_{(0)}$. We will
  consider it later.}
$\hat{C}_{(1)},\hat{C}_{(3)},\hat{\overline{C}}_{(1)},
\ \hat{\overline{C}}_{(3)}$
that may couple to the tachyon as in the Type~0B case.  Whatever the
couplings to the tachyon are, we can always diagonalize the kinetic
terms. We denote the potentials in the diagonal basis by
$\hat{C}^{+}_{(1)},\hat{C}^{+}_{(3)},\hat{C}^{-}_{(1)},\hat{C}^{-}_{(3)}$
but we will not make any assumption about the relation with the
original potentials. It is now evident that the fields with index $+$
(resp.~$-$) will couple to the tachyon through $f_{+}(\hat{T})$
(resp.~$f_{-}(\hat{T})$), since otherwise it would be impossible to get the
reduced action Eq.~(\ref{eq:0Baction9d-1}).

Thus, to the order considered, the Type~0A string effective action
must be of the form

\begin{equation}
\label{eq:0Aaction-1}
\begin{array}{rcl}
\hat{S}_{0A} & = & {\displaystyle\int} d^{10}\hat{x}\sqrt{|\hat{g}|}\,
\left\{e^{-2\hat{\phi}} \left[ \hat{R} -4(\partial\hat{\phi})^{2}
+\frac{1}{2\cdot 3!}\hat{H}^{2}+\frac{1}{2}(\partial\hat{T})^{2}
-V(\hat{T})\right] \right.\\
& & \\
& & 
\hspace{2cm}
\left.
+\sum_{\alpha=+,-}f_{\alpha}(\hat{T})
\left[-\frac{1}{4}\left(\hat{G}_{(2)}^{\alpha}\right)^{2} 
-\frac{1}{2\cdot 4!}\left(\hat{G}_{(4)}^{\alpha}\right)^{2}
\right]
\right\}\, ,\\
\end{array}
\end{equation}

\noindent where the field strengths are defined as in 
Eq.~(\ref{eq:10dfieldstrengths}) and the tachyon potential and 
coupling functions are identical to those of the Type~0B theory.

Let us now reduce this action to 9-dimensions in the direction of the
coordinate $x$. The relation between the 10-dimensional fields

\begin{equation}
\{\hat{g},
\hat{B},\hat{\phi},\hat{T}, 
\hat{C}^{+}_{(1)},\hat{C}^{+}_{(3)},
\hat{C}^{-}_{(1)},\hat{C}^{-}_{(3)}\}\, ,
\end{equation}

\noindent and the 9-dimensional fields 

\begin{equation}
\{g,B,A^{(1)},A^{(2)},k,\phi,T,
C^{+}_{(0)},C^{+}_{(1)},C^{+}_{(2)},C^{+}_{(3)},
C^{-}_{(0)},C^{-}_{(1)},C^{-}_{(2)},C^{-}_{(3)}\}\, ,  
\end{equation}

\noindent is, in the NSNS sector\footnote{We use the T-dual KK Ansatz. This
ensures that the resulting 9-dimensional actions are the same in stead
of being related by $k\rightarrow k^{-1}$ and 
$A^{(1)}\leftrightarrow A^{(2)}$.}

\begin{equation}
\begin{array}{rclrcl}
\hat{g}_{\mu\nu} & = & g_{\mu\nu}-k^{2}A^{(1)}{}_{\mu}A^{(1)}{}_{\nu}\, ,
\hspace{1cm}&
\hat{B}_{\mu\nu} & = & B_{\mu\nu}-A^{(1)}{}_{[\mu}A^{(2)}{}_{\nu]}\, ,
\\
& & & & &
\\
\hat{g}_{\mu\underline{x}} & = & -k^{2}A^{(1)}{}_{\mu}\, , &
\hat{B}_{\mu\underline{x}} & = & A^{(2)}{}_{\mu}\, ,
\\
& & & & &
\\
\hat{g}_{\underline{xx}} & = & -k^{2}\, , &
\hat{\phi}               & = & \phi +\frac{1}{2}\log{k}\, ,
\\
& & & & &
\\
\hat{T} & = & T\, .& & & \\
\end{array}
\label{eq:9dNSNS0A}
\end{equation}

\noindent This will give a 9-dimensional NSNS sector identical to that
of the action Eq.~(\ref{eq:0Baction9d-1}). The only possible relation
between the 10- and 9-dimensional RR fields is

\begin{equation}
\left\{
\begin{array}{rcl}
\hat{C}^{\pm}_{(2n-1)\, \mu_{1}\cdots\mu_{2n-1}}
& = & 
C^{\pm}_{(2n-1)\, \mu_{1}\cdots\mu_{2n-1}}
+(2n-1) A^{(1)}{}_{[\mu_{1}}
C^{\pm}_{(2n-2)\, \mu_{2}\cdots\mu_{2n-1}]}\, , \\
& & \\
\hat{C}^{\pm}_{(2n-1)\, \mu_{1}\cdots\mu_{2n-2}\underline{x}}
& = & 
C^{\pm}_{(2n-2)\, \mu_{1}\cdots\mu_{2n-2}}\, ,\\
\end{array}
\right.
\label{eq:9dRR0A-1}
\end{equation}

\noindent and we remark that it involves $A^{(1)}$ and not 
$A^{(2)}$, as in the Type~0B case. We cannot change this without 
spoiling T-duality in the NSNS sector.

The field strengths are related in flat indices by

\begin{equation}
\left\{
\begin{array}{rcl}
\hat{H}_{abc} & = & H_{abc}\, ,\\
& & \\
\hat{H}_{abx} & = & k^{-1}F^{(2)}{}_{ab}\, ,\\
\end{array}
\right.
\end{equation}

\noindent in the NSNS sector where the 9-dimensional field strengths
are also given by Eq.~(\ref{eq:d9NSNSfieldstrengths}). 

The RR field strengths are related by

\begin{equation}
\left\{
\begin{array}{rcl}
\hat{G}^{\pm}_{(2n)\, a_{1}\cdots a_{2n}} & = & 
G^{\pm}_{(2n)\, a_{1}\cdots a_{2n}}\, ,\\
& & \\
\hat{G}^{\pm}_{(2n)\, a_{1}\cdots a_{2n-1}x} & = & 
k^{-1}G^{\pm}_{(2n-1)\, a_{1}\cdots a_{2n-1}}\, ,\\
\end{array}
\right.
\end{equation}

\noindent where the 9-dimensional ones  are defined as follows

\begin{equation}
\label{eq:9dRRfieldstrengths0A-1}
\left\{
\begin{array}{lcl}
G_{(2n+1)}^{\pm} & = & dC_{(2n)}^{\pm} \, ,\\
& & \\
G_{(2n)}^{\pm} & = & dC_{(2n-1)}^{\pm} +F^{(1)} C_{(2n-2)}^{\pm}\, .\\
\end{array}
\right.
\end{equation}

\noindent The even ones involve $F^{(1)}$ while in the Type~0B
the odd ones involve $F^{(2)}$.

Summarizing, we have obtained the action

\begin{equation}
\begin{array}{rcl}
S & = & {\displaystyle\int} d^{9}x\,  \sqrt{|g|}\ 
\left\{ e^{-2\phi}
\left[R -4(\partial\phi)^{2} +{\textstyle\frac{1}{2\cdot 3!}}H^{2}
+(\partial\log{k})^{2} 
\right.
\right.\\
& & \\
& &
\hspace{3.5cm}
\left.
-{\textstyle\frac{1}{4}}k^{2} \left(F^{(1)}\right)^{2} 
-{\textstyle\frac{1}{4}}k^{-2}\left(F^{(2)}\right)^{2}
+\textstyle{\frac{1}{2}}\left(\partial T\right)^{2} -V(T)
\right]
\\
& & \\
& & 
\hspace{2cm}
+\sum_{\alpha=+,-}f_{\alpha}(T)
\left[\frac{1}{2}k^{-1}\left(G_{(1)}^{\alpha}\right)^{2} 
-\frac{1}{4}k\left(G_{(2)}^{\alpha}\right)^{2} 
\right.
\\
& & \\
& & 
\hspace{5cm}
\left.
\left.
+\frac{1}{2\cdot 3!}k^{-1}\left(G_{(3)}^{\alpha}\right)^{2}
-\frac{1}{2\cdot 4!}k\left(G_{(4)}^{\alpha}\right)^{2}
\right]
\right\}\, .\\
\end{array}
\label{eq:0Aaction9d-1}
\end{equation}

This action is different from the one we obtained from the Type~0B
theory Eq.~(\ref{eq:0Baction9d-1}) in two points: the definition of
the 9-dimensional field strengths involves only one of the two
9-dimensional vectors: the Kaluza-Klein one. Since they are
interchanged by T-duality, we need both vectors to appear in the 
field strengths.
On the other hand, in the Type~0B case we have obtained one
RR field strength which is not present in the reduced Type~0A theory:
$G^{+}_{(5)}$ and in the Type~0A case we have obtained another RR
field strength absent in the reduced Type~0B action: $G^{-}_{(4)}$.

The first problem can only be solved by making the winding vector
appear in the reduced RR field strengths, which implies that $\hat{B}$
must appear in the 10-dimensional RR field strengths. Up to possible
field redefinitions, there is only one way of doing this: precisely
defining the RR field strengths as in the Type~II theories, i.e.

\begin{equation}
\label{eq:RRfieldstrengths}
\hat{G}^{\pm}_{(n)} = 
d\hat{C}^{\pm}_{(n-1)} -\hat{\cal H}\hat{C}^{\pm}_{(n-3)}\, ,
\end{equation}

\noindent in both the Type~0B and 0A theories. This is consistent 
with the fact that the amplitudes involving two RR fields of the same
sector and a NSNS field (different fro the tachyon) are identical to
those of the Type~IIB$_{\pm}$ theories. The only difference could be
the sign of the second term. We can set it to minus, as above, for the
Type~0B $\hat{G}^{+}_{(2n+1)}$ field strengths by fixing the relative
sign between $\hat{\cal B}$ and the $\hat{C}^{+}_{(2n)}$ potentials.
In principle the sign of the second term in the Type~0B field
strengths $\hat{G}^{-}_{(2n+1)}$ could still be arbitrarily chosen by
changing the sign of all the RR potentials $\hat{C}^{-}_{2n}$ because
the $+$ and $-$ RR potentials are decoupled in the action
Eq.~(\ref{eq:0Baction9d-1}). However, as we are going to argue next,
we are going to have to introduce a Chern-Simons term that may couple
them and we have to be open to the two possible signs.

The second problem can only be solved by Hodge-dualizing $G^{+}_{(5)}$
and identifying the dual field with $-G^{-}_{(4)}$.  This is somewhat
reminiscent of the procedure followed in Type~II theories
\cite{kn:MO}. For this dualization to give the right form of
$G^{-}_{(4)}$ it will be necessary to add to the 10-dimensional
Type~0B action a Chern-Simons term and this will force us to introduce
another one in the 10-dimensional Type~0A action. A subtle point
arises here: when one dualizes a field strength whose kinetic term
comes multiplied by a function, the kinetic term of the dual field
comes multiplied by the {\it inverse} function. Thus, the $G^{-}_{(4)}$
kinetic term will carry an $f_{+}^{-1}(T)k$ factor. We expected the
$k$ factor, but we also expected an $f_{-}(T)$ factor. To establish
T-duality, then, we must have
$f^{-1}_{+} =f_{-}$
%\begin{equation}
%f^{-1}_{+}(T) =f_{-}(T)\, ,
%\end{equation}
%%\noindent This constrains the possible corrections to these functions.
which together with  Eq.~(\ref{eq:relatie1}) implies that
\begin{equation}
  \label{eq:def_f}
  f_{\pm}\left(\hat{\mathcal{T}}\right) \;=\; 
       \exp\left( \pm h(\hat{\mathcal{T}})\right) \; ,
\end{equation}
where $h$ is an odd function of $\hat{\mathcal{T}}$.
Note that this result was anticipated in Ref. \cite{kn:KlTs3}
by means of tadpole considerations; here it arises as a necessity
for T-duality to work. 
\par
The need to introduce a 10-dimensional Chern-Simons term can also be
seen directly in 10-dimensions starting with the NSD action with the
kinetic terms Eq.~(\ref{eq:NSDkineticterms}). It is instructive to
derive the Chern-Simons term using an argument different from
T-duality. Let us for the moment set to zero the tachyon field, in order 
to simplify the calculations (in any case, it does not play any role in
the determination of the Chern-Simons term). The kinetic terms are
just

\begin{equation}
\int d^{10}\hat{x}\sqrt{|\hat{\jmath}|}\,
\left[ 
{\textstyle\frac{1}{2\cdot 5!}}\left(\hat{G}_{(5)}\right)^{2}
+{\textstyle\frac{1}{2\cdot 5!}}\left(\hat{\overline{G}}_{(5)}\right)^{2}
\right]\, .
\end{equation}

The Chern-Simons term has to be such that selfduality of
$\hat{G}_{(5)}$ and the anti-selfduality of $\hat{\overline{G}}_{(5)}$ can
be consistently imposed, i.e.~such that the equations of motion are
identical to the Bianchi identities, 

\begin{equation}
\begin{array}{rcl}
d\left(\hat{G}_{(5)} +\hat{H}\hat{C}_{(2)}\right) 
& = & 0\, ,\\
& & \\
d\left(\hat{\overline{G}}_{(5)} +\hat{H}\hat{\overline{C}}_{(2)}\right) 
& = & 0\, ,\\
\end{array}
\end{equation}

\noindent using the (anti-) selfduality constraints. The Chern-Simons 
is, therefore, given by the addition of the Type~IIB$_{+}$ and
Type~IIB$_{-}$ Chern-Simons terms, namely

\begin{equation}
\int d^{10}\hat{x}\sqrt{|\hat{\jmath}|}\, \left\{
{\textstyle\frac{1}{2\cdot 5!}}\left(\hat{G}_{(5)}\right)^{2}
+{\textstyle\frac{1}{2\cdot 5!}}\left(\hat{\overline{G}}_{(5)}\right)^{2}
+{\textstyle\frac{10}{(5!)^{2}}}\hat{\epsilon}
\left[\hat{G}_{(5)}\hat{\cal H}\hat{C}_{(2)}
-\hat{\overline{G}}_{(5)}\hat{\cal H}\hat{\overline{C}}_{(2)}\right]
\right\}\, ,
\end{equation}

\noindent which, written in terms of the diagonal fields is

\begin{equation}
\int d^{10}\hat{x}\sqrt{|\hat{\jmath}|}\, \left\{
{\textstyle\frac{1}{4\cdot 5!}}\left(\hat{G}^{+}_{(5)}\right)^{2}
+{\textstyle\frac{1}{4\cdot 5!}}\left(\hat{G}^{-}_{(5)}\right)^{2}
+{\textstyle\frac{1}{4!\cdot5!}}\hat{\epsilon}
\left[\hat{G}^{+}_{(5)}\hat{\cal H}\hat{C}^{-}_{(2)}
+\hat{G}^{-}_{(5)}\hat{\cal H}\hat{C}^{+}_{(2)}\right]
\right\}\, .
\end{equation}

We can now Poincar\'e-dualize $\hat{G}^{-}_{(5)}$, adding to the above
action a Lagrange-multiplier term to enforce its Bianchi identity

\begin{equation}
\int d^{10}\hat{x}\sqrt{|\hat{\jmath}|}\, {\textstyle\frac{1}{4!\cdot 5!}}
\hat{\epsilon}\partial\tilde{\hat{C}}{}^{-}_{(4)}
\left(\hat{G}_{(5)} +10\hat{H}\hat{C}_{(2)}\right)\, ,
\end{equation}

\noindent then, solving for $\hat{G}^{-}_{(5)}$ 

\begin{equation}
\hat{G}^{-}_{(5)}= {}^{\star}\tilde{\hat{G}}{}^{-}_{(5)}\, ,
\hspace{1cm}
\tilde{\hat{G}}{}^{-}_{(5)} =5\partial\tilde{\hat{C}}{}^{-}_{(4)}
-10\hat{\cal H}\hat{C}^{-}_{(2)}\, ,
\end{equation}

\noindent and substituting this solution into the action and, finally,
identifying $\tilde{\hat{G}}{}^{-}_{(5)}=\hat{G}{}^{+}_{(5)}$,
we find

\begin{equation}
\label{eq:CS0B-1}
\int d^{10}\hat{x}\sqrt{|\hat{\jmath}|}\, \left[
{\textstyle\frac{1}{2\cdot 5!}}\left(\hat{G}^{+}_{(5)}\right)^{2}
-{\textstyle\frac{1}{4 \cdot5!}}\frac{\hat{\epsilon}}{\sqrt{|\hat{\jmath}|}}
\hat{G}^{+}_{(5)}\partial\hat{C}^{-}_{(2)}\mathcal{B}
\right]\, ,
\end{equation}

\noindent which contains the actual kinetic term that we have and
the Chern-Simons term that we should expect.

%%%%%%%%%%%%%%%%%%%%%%%%%%%%%%%%%%%%%%%%%%%%%%%%%%%%%%%%%%%%%%%%%%%%%%

\subsection{Corrected 10-dimensional Type~0A/B Effective Actions
and T-Duality Rules}

It is quite straightforward to carry on with the program. First, we
consider again the action Eq.~(\ref{eq:0Baction-1}) but with RR field
strengths given by Eq.~(\ref{eq:RRfieldstrengths}) in both, $+$ and
$-$, sectors and repeat the dimensional reduction. The Kaluza-Klein
Ansatz is the same for all the fields, and the 10-dimensional field
strengths decompose in 9-dimensional field strengths in the same form
and so, we get an action of the form Eq.~(\ref{eq:0Baction9d-1}), but
with lower-dimensional RR-field strengths defined by

\begin{equation}
\label{eq:9dRRfieldstrengths0AB}
\left\{
\begin{array}{lcl}
G^{\pm}_{(2n+1)} & = & dC^{\pm}_{(2n)} -HC^{\pm}_{(2n-2)} 
+F^{(2)} C^{\pm}_{(2n-1)}\, ,\\
& & \\
G^{\pm}_{(2n)} & = & dC^{\pm}_{(2n-1)} -HC^{\pm}_{(2n-3)} 
+F^{(1)} C^{\pm}_{(2n-2)}\, .\\
\end{array}
\right.
\end{equation}

\noindent Now, both the Kaluza-Klein and winding vector field are
present in the 9-dimensional RR field strengths.

The next step is to Poincar\'e-dualize $G^{+}_{(5)}$ into $G^{-}_{(4)}$:
we add to the 9-dimensional action a Lagrange-multiplier term to enforce the 
Bianchi identity

\begin{equation}
d\left[G^{+}_{(5)}+HC^{+}_{(2)}-F^{(2)}C^{+}_{(3)}\right]=0\, .  
\end{equation}

The Lagrange multiplier has to be a 3-form that will become the dual
potential $C^{-}_{(3)}$. Then, the Lagrange multiplier term will take
the form

\begin{equation}
\label{eq:lagragemultiplierterm}
\alpha\int d^{9}x\, \epsilon\, \partial C^{-}_{(3)} 
\left[G^{+}_{(5)}+10HC^{+}_{(2)}-10F^{(2)}C^{+}_{(3)}\right]\, ,
\end{equation}

\noindent where $\alpha$ is a constant whose value has to be chosen 
so as to get the right normalization for the kinetic term of
$C^{-}_{(3)}$. In the action Eq.~(\ref{eq:0Baction9d-1}) with the
above Lagrange-multiplier term, $C^{+}_{(4)}$ only appears through
$G^{+}_{(5)}$. We can consider it as a functional of $G^{+}_{(5)}$
since we can always recover the expression of $G^{+}_{(5)}$ in terms of 
$C^{+}_{(4)}$ through equation of motion of the Lagrange-multiplier.
Now, the $G^{+}_{(5)}$ equation of motion is

\begin{equation}
G^{+}_{(5)}= -\alpha f^{-1}_{+}(T) k\frac{\epsilon}{\sqrt{|g|}} 
\partial C^{-}_{(3)}\, .
\end{equation}

\noindent We expected 

\begin{equation}
\label{eq:expected}
G^{+}_{(5)}=  -\sqrt{f_{-}/f_{+}}\, k\, {}^{\star}G^{-}_{(4)}\, ,
\end{equation}

\noindent with 

\begin{equation}
G^{-}_{(4)} = dC^{-}_{(3)} -HC^{-}_{(1)} +F^{(1)} C^{-}_{(2)}\, .
\end{equation}

\noindent This fixes the normalization constant 
$\alpha=+\frac{1}{3!\cdot5!}$, implies $f^{-1}_{+}(T) =f_{-}(T)$ and
also tells us that there should be a 9-dimensional Chern-Simons term
in the 9-dimensional action Eq.~(\ref{eq:0Baction9d-1}) of the form

\begin{equation}
-{\textstyle\frac{1}{2\cdot 3!\cdot5!}}\int d^{9}x\, \epsilon\, G^{+}_{(5)}
\left[2HC^{-}_{(1)}-3F^{(1)}C^{-}_{(2)}\right]\, ,
\end{equation}

\noindent to get Eq.~(\ref{eq:expected}). This term can only come
from the 10-dimensional Chern-Simons term in
Eq.~(\ref{eq:CS0B-1}) that we can also write, up to total derivatives,
in the form

\begin{equation}
\label{eq:CS0B-2}
-{\textstyle\frac{1}{96}} \int d^{10}\hat{x}\, 
\hat{\epsilon}\,
\partial\hat{C}^{+}_{(4)}\partial\hat{C}^{-}_{(2)}\hat{\cal B}\, ,
\end{equation}

\noindent which gives rise to the term we wanted and another term 
not involving $C^{+}_{(4)}$ in any way:

\begin{equation}
-{\textstyle\frac{1}{2\cdot 3!\cdot 5!}} \int d^{9}x\,
\epsilon\,
\left[G^{+}_{(5)}\left(2HC^{-}_{(1)} -3F^{(1)}C^{-}_{(2)}\right) 
-5G^{+}_{(4)}HC^{-}_{(2)}
\right]\, .
\end{equation}

Observe that the Chern-Simons term
Eq.~(\ref{eq:CS0B-1},\ref{eq:CS0B-1}) is very similar to the
Chern-Simons term in the Type~IIB NSD string effective action
\cite{kn:BBO,kn:MO}. Here, however, it mixes non-trivially the 
two RR sectors.

Adding to this 9-dimensional Chern-Simons the Lagrange-multiplier term
Eq.~(\ref{eq:lagragemultiplierterm}) with the value of $\alpha$ that
we have calculated, we find the equation of motion
Eq.~(\ref{eq:expected}), and, using it to eliminate $G^{+}_{(5)}$, we
finally get the 9-dimensional Type~0 string effective action

\begin{equation}
\begin{array}{rcl}
S & = & {\displaystyle\int} d^{9}x\,  \sqrt{|g|}\ 
\left\{ e^{-2\phi}
\left[R -4(\partial\phi)^{2} +{\textstyle\frac{1}{2\cdot 3!}}H^{2}
+(\partial\log{k})^{2} 
\right.
\right.\\
& & \\
& &
\hspace{3.5cm}
\left.
-{\textstyle\frac{1}{4}}k^{2} \left(F^{(1)}\right)^{2} 
-{\textstyle\frac{1}{4}}k^{-2}\left(F^{(2)}\right)^{2}
+\textstyle{\frac{1}{2}}\left(\partial T\right)^{2} -V(T)
\right]
\\
& & \\
& & 
\hspace{2cm}
+\sum_{\alpha=+,-}f_{\alpha}(T)
\left[\frac{1}{2}k^{-1}\left(G_{(1)}^{\alpha}\right)^{2} 
-\frac{1}{4}k\left(G_{(2)}^{\alpha}\right)^{2} 
\right.
\\
& & \\
& & 
\hspace{5cm}
\left.
+\frac{1}{2\cdot 3!}k^{-1}\left(G_{(3)}^{\alpha}\right)^{2}
-\frac{1}{2\cdot 4!}k\left(G_{(4)}^{\alpha}\right)^{2}
\right]
\\
& & \\
& & 
\hspace{1cm}
-\frac{1}{36}\frac{\epsilon}{\sqrt{|g|}} 
\left\{ 
\partial C^{+}_{(3)}\partial C^{-}_{(3)} A^{(2)}
-\frac{9}{2} C^{+}_{(2)}C^{-}_{(2)} \partial A^{(1)}
\left(\partial B -A^{(1)}\partial A^{(2)} -A^{(2)}\partial A^{(1)}\right)
\right. 
\\
& & \\
& & 
\hspace{2cm}
+\frac{3}{2}
\left[
\partial C^{+}_{(3)}\partial C^{-}_{(2)} \left(B+A^{(1)}A^{(2)}\right)
+2\partial C^{+}_{(3)}C^{-}_{(2)} A^{(2)} \partial A^{(1)} A^{(2)}
\right.
\\
& & \\
& & 
\hspace{2.5cm}
\left.
\left.
\left.
+\partial C^{-}_{(3)}\partial C^{+}_{(2)} \left(B+A^{(1)}A^{(2)}\right)
+2\partial C^{-}_{(3)}C^{+}_{(2)} A^{(2)} \partial A^{(1)} A^{(2)}
\right]
\right\}
\right\}\, .\\
\end{array}
\label{eq:0action9d}
\end{equation}

In order to establish T-duality then, we have to find a
10-dimensional Chern-Simons term to add to the
Type~0A string effective action Eq.~(\ref{eq:0Aaction9d-1}) leading to
the above 9-dimensional Type~0 string effective action using the same
Ansatz as before. This is a very non-trivial check of our construction.
It takes little time to see that the sought for Chern-Simons term is

\begin{equation}
\label{eq:CS0A-1}
-{\textstyle\frac{1}{72}} \int d^{10}\hat{x}\, 
\hat{\epsilon}\,
\partial\hat{C}^{+}_{(3)}\partial\hat{C}^{-}_{(3)}\hat{B}\, .
\end{equation}

\noindent Again, this Chern-Simons term looks very similar to the one in the
Type~IIA string effective action. In fact, we could rewrite it in the
form

\begin{equation}
\label{eq:CS0A-2}
-{\textstyle\frac{1}{144}} \int d^{10}\hat{x}\, 
\hat{\epsilon}\,\left[
\partial\hat{C}_{(3)}\partial\hat{C}_{(3)}\hat{B}
-\partial\hat{\overline{C}}_{(3)}\partial\hat{\overline{C}}_{(3)}\hat{B}
\right]\, .
\end{equation}

\noindent which would be the sum of the Chern-Simons terms of the 
Type~IIA$_{+}$ and Type~IIA$_{-}$ theories (which are related
by target space parity).  

The resulting Type~0A string effective action
(Eq.~(\ref{eq:0Aaction-1}) plus the Chern-Simons term
Eq.~(\ref{eq:CS0A-1})) is left-right invariant (i.e.~invariant under
the interchange of the two RR sectors $\hat{C}^{\pm}\rightarrow \pm
\hat{C}^{\pm}$ and sign reversal of the Kalb-Ramond form $\hat{B}
\rightarrow -\hat{B}$), as it should.

In the same way, the complete Type~0B action
(Eq.~(\ref{eq:0Baction-1}) plus Eq.~(\ref{eq:CS0B-2})) is invariant
under the transformation that changes the sign of the tachyon and
interchanges the $+$ and $-$ RR field strengths if we dualize
$\hat{G}^{-}_{(5)}$ into $\hat{G}^{+}_{(5)}$.

We have just established T-duality between the Type~0A and~0B string
effective actions, as we intended to do. The T-duality rules are
identical to those of the Type~II theories \cite{kn:MO}, but now
working inside each of the ${}^{+}$ and ${}^{-}$ diagonal RR sectors.

\section{Democratic Type~0 Actions and Massive 0A}
\label{sec:DemAction}
In Ref.~\cite{kn:BKORvP} a ``democratic'' pseudo-action for Type~II
theories was proposed in which all RR potentials appear on the same
footing. The pseudo-action has to be supplemented by duality
constraints relating ``electric'' and ``magnetic'' RR fields (hence
the ``pseudo'') and one of its properties is that it has no
Chern-Simons term and only the kinetic terms for all the field
strengths appear in it. In the Type~0 case, it is a simple exercise to
get an action in which RR field strengths of all orders appear in the
same footing: in the 0B action we can dualize $\hat{G}^{-}_{(3)}$ and
$\hat{G}^{-}_{(1)}$ into $\hat{G}^{+}_{(7)}$ and $\hat{G}^{+}_{(9)}$
respectively, and in this order\footnote{We cannot directly dualize
  $\hat{G}^{-}_{(1)}$ because there are explicit $\hat{C}^{-}_{(0)}$
  potentials in $\hat{G}^{-}_{(3)}$. We could absorb them into a
  redefinition of $\hat{C}^{-}_{(2)}$, but this would introduce
  unnecessary complications.} by the standard Poincar\'e-dualization
procedure. There is no need to impose any duality constraint as the
resulting $\hat{G}^{+}_{(2n+1)}\, ,\,\,\,\, n=0,1,2,3,4$ field
strengths are independent. Actually, not all ``electric'' and
``magnetic'' field strengths appear, but only some electric and some
magnetic. In any case, the action obtained in this way is really much
simpler than the one we arrived at in the previous section
given by Eq.~(\ref{eq:0Baction-1}) plus
Eq.~(\ref{eq:CS0B-1}) or Eq.~(\ref{eq:CS0B-2}) with RR field strengths
given by Eq.~(\ref{eq:RRfieldstrengths}). In particular, there is no
Chern-Simons term and only the kinetic terms of all field strengths
$\hat{G}^{+}_{(2n+1)}$ $(n=0,1,2,3,4$) appear:

\begin{equation}
\label{eq:0Baction-2}
\begin{array}{rcl}
\hat{S}_{0B} & = & {\displaystyle\int} d^{10}\hat{x}\sqrt{|\hat{\jmath}|}\,
\left\{e^{-2\hat{\varphi}} \left[ \hat{R} -4(\partial\hat{\varphi})^{2}
+\frac{1}{2\cdot 3!}\hat{\cal H}^{2}+\frac{1}{2}(\partial\hat{\cal T})^{2}
-V(\hat{\cal T})
\right] \right.\\
& & \\
& & 
\hspace{2.5cm}
\left.
+f_{+}(\hat{\cal T})\sum_{n=1}^{n=4}\frac{1}{2\cdot (2n+1)!}
\left(\hat{G}_{(2n+1)}^{+}\right)^{2} 
\right\}\, .\\
\end{array}
\end{equation}

We could have dualized instead the $\hat{G}^{+}_{(5)}$,
$\hat{G}^{+}_{(3)}$ and $\hat{G}^{+}_{(1)}$ field strengths, in this
order, and we would have obtained the above action with the $+$ indices
replaced by $-$ indices. The transformation that changes the
sign of the tachyon and interchanges the two RR sectors would take us
back to the above action which is, thus, invariant under a combination
of that transformation and the Poincar\'e-dualization of all the field
strengths.

Needless to say one can also write down a democratic
formulation of the Type~0A
action created in the foregoing section and it has the same features
as the democratic 0B action, namely only kinetic terms appear.
T-duality is then established by extending the decomposition
rules (\ref{eq:9dRR0B-1},\ref{eq:9dRR0A-1}) to include the higher
(``magnetic'') RR forms.

In Ref. \cite{kn:BRGPT} it was shown that in order to establish
T-duality between Type~IIB and massive Type~IIA, Romans' theory
for short \cite{kn:Ro2}, one has to apply 
{\em Generalized Dimensional Reduction}
on the Type~IIB side and standard dimensional reduction on Romans' side.
The symmetry abused to perform the GDR is the invariance under
the addition of constants to the Type~IIB RR scalar,
{\em i.e.} $\delta \hat{C}_{(0)}= m= cte.$ 

In the democratic formulation, the symmetry under constant shifts
of the two RR zero forms also acts on the higher RR forms
and can be written as
\begin{equation}
  \label{eq:shiftsymmetry-2}
  \hat{C}^{\pm}\;=\; \hat{C}^{\pm}\;+\; a^{\pm}e^{\hat{B}} \; ,
\end{equation}

so we can apply GDR in much the same way as in Ref. \cite{kn:BRGPT}
and oxidize the 9-dimensional theory to the massive 0A action.
In the democratic 0B action there
is however only one RR scalar present so that it might seem that 
we would end up with only one mass parameter, whereas Type~0A can
support 2 mass parameters associated to the 2 D8-branes present in
its spectrum. This is however only an illusion: the 9-form field strength
in Type~0B will induce a 9-form field strength in d=9, which in its 
turn can only be related to a 10-form field strength in Type~0A. It
is this 10-form field strength that couples to the second D8-brane.

Generalized Dimensional Reduction, then, boils down to using the
decomposition\footnote{Please note the $\hat{}$ on the $B$.}
\begin{equation}
  \label{eq:9dRR0B-mas}
  \hat{C}_{(2n)}^{+}\;=\; C_{(2n)}^{+}\,-\, C^{+}_{(2n-1)}\left(
                                                    d\underline{y}+A^{(2)}
                                                  \right) 
                          \,+\, \underline{y}\  G_{(0)}^{+}
                          \textstyle{\frac{1}{n!}}\hat{B}^{n}\; ,
\end{equation}
in stead of (\ref{eq:9dRR0B-1}), in the 
reduction carried out in Sec.~(\ref{sec:type0-9d}).
The resulting 9-dimensional action can be obtained by standard
dimensional reduction from the action

\begin{equation}
\label{eq:0Aaction-2}
\begin{array}{rcl}
\hat{S}_{0A} & = & {\displaystyle\int} d^{10}\hat{x}\sqrt{|\hat{g}|}\,
\left\{e^{-2\hat{\phi}} \left[ \hat{R} -4(\partial\hat{\phi})^{2}
+\frac{1}{2\cdot 3!}\hat{H}^{2}+\frac{1}{2}(\partial\hat{\cal T})^{2}
-V(\hat{\cal T})\right] \right.\\
& & \\
& & 
\hspace{1cm}
\left.
-f_{+}(\hat{\cal T})\sum_{n=0}^{n=5}\frac{1}{2\cdot (2n)!}
\left(\hat{G}_{(2n)}^{+}\right)^{2} 
\right\}\, .\\
\end{array}
\end{equation}
where the RR field strengths read
\begin{equation}
  \hat{G}^{\pm}\;=\; d\hat{C}^{\pm}\,-\,\hat{H}\hat{C}^{\pm}\,+\,
                     G_{(0)}^{\pm}e^{B}\; .
\end{equation}

The non-democratic formulation of massive 0A can be obtained by
dualizing the 10-, 8- and 6-form field strengths, resulting in

\begin{equation}
\label{eq:CS0A-3}
\begin{array}{rcl}
\hat{S}_{0A} & = & {\displaystyle\int} d^{10}\hat{x}\sqrt{|\hat{g}|}\,
\left\{e^{-2\hat{\phi}} \left[ \hat{R} -4(\partial\hat{\phi})^{2}
+\frac{1}{2\cdot 3!}\hat{H}^{2}+\frac{1}{2}(\partial\hat{\cal T})^{2}
-V(\hat{\cal T})\right]\right.
\\
& & \\
& & 
\hspace{2cm}
-\sum_{\alpha =\pm} f_{\alpha}
     \sum_{n=0}^{2} \textstyle{\frac{1}{2\cdot (2n)!}} G_{(2n)}^{\alpha\ 2} 
\\
& & \\
& &
-\left. {\textstyle\frac{1}{72\sqrt{\hat{g}}}} 
\hat{\epsilon}\,
\left[
\partial\hat{C}^{+}_{(3)}\partial\hat{C}^{-}_{(3)}\hat{B}
+{\textstyle\frac{1}{4}}\hat{G}^{-}_{(0)}\partial\hat{C}^{+}_{(3)}
\hat{B}^{3}
+{\textstyle\frac{1}{4}}\hat{G}^{+}_{(0)}\partial\hat{C}^{-}_{(3)}
\hat{B}^{3}
+{\textstyle\frac{9}{80}}\hat{G}^{+}_{(0)}\hat{G}^{-}_{(0)}
\hat{B}^{5}
\right]\right\}
\, ,
\end{array}
\end{equation}

\noindent which is just what one would expect.

%%%%%%%%%%%%%%%%%%%%%%%%%%%%%%%%%%%%%%%%%%%%%%%%%%%%%%%%%%%%%%%%%%%%%%
\section{Type~0 D-brane solutions from Type~II}
%%%%%%%%%%%%%%%%%%%%%%%%%%%%%%%%%%%%%%%%%%%%%%%%%%%%%%%%%%%%%%%%%%%%%%
In this section we are going to see how to adapt Type~II BPS solutions
to the Type~0 setting. This will be done under the assumption of a
constant tachyon field $\hat{\cal T}_{0}$, and we will absorb any
tachyon dependence in the equations of motion into the RR-fields.  In
order to do this consistently however, we must investigate the tachyon
equation of motion, {\em i.e.}
\begin{equation}
  \label{eq:tacheom}
  \nabla^{\mu}\left( e^{-2\phi}\partial_{\mu}\mathcal{T}\right) 
+V^{\prime}(\hat{\cal T})
    - h^{\prime}(\mathcal{T})\left[
          \sum_{n} \textstyle{\frac{(-)^{n}}{2\cdot n!}}F^{+\ 2}_{(n)}\,-\,
          \sum_{n} \textstyle{\frac{(-)^{n}}{2\cdot n!}}F^{-\ 2}_{(n)} 
     \right] =0\, ,
\end{equation}
where we made use of Eq.~(\ref{eq:def_f}) and $F^{\pm}_{(n)}$ denotes
the rescaled field strength.  It is natural to choose a constant value
of the tachyon $\hat{\cal T}_{0}$ that minimizes the potential
$V^{\prime}(\hat{\cal T}_{0})=0$. We will assume that a minimum exists
and, further, that $h^{\prime}(\hat{\cal T}_{0})$ is
finite\footnote{If $h^{\prime}(\hat{\cal T}_{0})=0$ we end up with no
  extra constraint and we can embed every Type~II solution in Type~0.
  Therefore, from now on we will consider the $h^{\prime}\neq 0$ case
  only.}. The tachyon equation of motion, then, leads to the
constraint
\begin{equation}
\label{eq:constraint}
  \sum_{n} \textstyle{\frac{(-)^{n}}{2\cdot n!}}F^{+\ 2}_{(n)}\,-\,
          \sum_{n} \textstyle{\frac{(-)^{n}}{2\cdot n!}}F^{-\ 2}_{(n)}=0\, . 
\end{equation}

In terms of the rescaled RR field strengths, denoted by $F$, the
equations of motions can be written as,\footnote{We will adapt the
  same philosophy as in \cite{kn:JMO} meaning that in the equations of
  motion also the dual fields are given. This for instance means that
  not only $F_{(5)}^{+}$, but also $F_{(5)}^{-}$ will contribute.}
\begin{eqnarray}
  \label{eq:0BEoM}
  0 &=& d{}^{\star}F^{\pm} +H\wedge F^{\pm}\, , \nonumber \\
  0 &=& d \left(e^{-2\phi} {}^{\star} H \right)
          +\frac{1}{2} \sum_{\alpha}
              {}^{\star} F^{\alpha} \wedge F^{\alpha}\, , \nonumber \\
  0&=& R +4\left(\partial\phi\right)^{2}  -4\nabla^{2}\phi
       +{\textstyle\frac{1}{2\cdot 3!}}H^{2}\, , \nonumber \\
  R_{\mu\nu}&=& 2\nabla_{\mu}\nabla_{\nu}\phi
          -{\textstyle\frac{1}{4}} H_{\mu}{}^{\rho\sigma} H_{\nu\rho\sigma}
          +{\textstyle\frac{1}{4}}e^{2\phi}\sum_{n,\alpha}
          {\textstyle\frac{(-1)^{n}}{n!}} T^{\alpha (n)}{}_{\mu\nu}\, ,
\end{eqnarray}

\noindent where $T^{\pm (n)}{}_{\mu\nu}$ are the energy-momentum tensor
of the RR fields

\begin{equation}
T^{\pm (n)}{}_{\mu\nu} = n\ F^{\pm}{}_{(n)\mu}{}^{\rho_{1}\cdots\rho_{n-1}}
F^{\pm}_{(n)\nu\rho_{1}\cdots\rho_{n-1}}
-{\textstyle\frac{1}{2}} g_{\mu\nu} F^{\pm}_{(n)}{}^{2}\, .
\end{equation}

\noindent The Type~II equations of motion can be obtained from these
by taking for example all the $^{-}$ RR-fields to vanish.

Now, a typical Type~II brane solution cannot, except for the D3-brane,
be a solution of the constraint Eq.~(\ref{eq:constraint}), and so the
best thing we can do is to distribute each Type~II D-brane charge
evenly over the $^{+}$ and $^{-}$ $(p+2)$-form field
strength\footnote{A similar idea was proposed in
  Ref.~\cite{Armoni:1999gc}.}. The constraint is, then, automatically
satisfied and the solution reads
\begin{eqnarray}
  \label{eq:0_D-braneSol}
  ds^{2} &=& H^{-1/2}\left( dt^{2}-d\vec{y}^{\, 2}_{(p)} \right)
               -H^{1/2}d\vec{x}_{(9-p)}^{\, 2} \; ,\nonumber \\
  e^{2\phi} &=& H^{\frac{p-3}{4}} \; ,\nonumber \\
  C^{+}_{ty_{1}\ldots y_{p}} &=& \pm C^{-}_{ty_{1}\ldots y_{p}} \;=\;
                                 \textstyle{\frac{1}{\sqrt{2}}}H^{-1} \; ,
\end{eqnarray}
where $H$ only depends on the transverse coordinates,
$\vec{x}_{(9-p)}$, and is harmonic.  Since this solution bears $^{+}$
and $^{-}$ charge, but has the form of only one object we are destined
to interpret these solutions as the $Dp_{\pm}$-brane \cite{kn:costa},
a bound state of a $Dp_{+}$- and a $Dp_{-}$-brane\footnote{Please note
  that in this notation the system of a coincident electric and
  magnetic $D3$ brane \cite{kn:KlTs3} is denoted by $D3_{\pm}$.}.
Observe that these solutions are simpler in terms of the original (but
rescaled), non-diagonal $C_{(p+1)},\overline{C}_{(p+1)}$ RR potentials
because they are charged only with respect to one of them.  They are
also trivially related by T-duality as Type~II D$p$-branes are.
\par
The fact that this pairing occurs follows naturally from the Type~IIB
$D3$-brane solution: Since it is selfdual it automatically satisfies
the condition (\ref{eq:constraint}), but as before the $D3$-brane
charge must be divided by $\sqrt{2}$ in order to satisfy the equations
of motion.  Consider then T-duality in a worldvolume direction; in the
democratic formulation, the electric component of the 5-form field
strength gives rise to the electric component of $G_{(4)}^{+}$,
whereas its magnetic part gives rise to a magnetic $G_{(6)}^{+}$ and
thus leads to an electric $G_{(4)}^{-}$. Needless to say it works also
in the other direction, implying that all the $Dp_{\pm}$ branes are
connected by T-duality.
\par
In Ref. \cite{kn:costa} it was shown that the potential between a
$D(p+r)_{\pm}$-brane and a $D(p+s)_{\pm}$ vanishes if $r+s=4$.  This
means that we can expect the, adapting the notation of
\cite{kn:BdREJvdS} to the case at hand, $\left( p\mid
  D(p+r)_{\pm},D(p+4-r)_{\pm}\right)$ intersection to be described by
the {\it harmonic superposition rule} \cite{kn:Ts-M}.  In Type~II the
$r+s=4$ class can be generated by T-duality from the $\left( 1\mid
  D3,D3\right)$, which in the Type~0 setting has to be interpreted as
a $\left( 1\mid D3_{\pm},D3_{\pm}\right)$ intersection.  Since we
embed a solution which is based on a selfdual 5-form, Eq.
(\ref{eq:tacheom}) is identically satisfied, and as before will give
rise to a solution once we divide the Type~II RR-field Ansatz by
$\sqrt{2}$.  Applying T-duality to this Type~0 intersection, we can
generate the whole class of $\left( p\mid
  D(p+r)_{\pm},D(p+4-r)_{\pm}\right)$ intersections.\footnote{Note
  that this will not reproduce \cite{kn:costa}'s $D5_{\pm}D1_{\pm}$
  solution as we took the tachyon potential to be zero in contrast to
  \cite{kn:costa}.}  Please observe that Eq.~(\ref{eq:tacheom}) is
automatically obeyed: by using the $Dp_{\pm}$-branes in stead of a
$Dp_{+}$-brane, say, we are effectively identifying the $^{+}$ and
$^{-}$ sector, automatically satisfying Eq.~(\ref{eq:constraint}).
\par
It should be clear that we can embed every Type~II solution into
Type~0: just distribute the Type~II RR-charge over the $^{+}$ and
$^{-}$ sector in the appropriate way.  {}For Type~II brane solutions
this means changing the $Dp$-brane for a $Dp_{\pm}$-brane.  In
particular this means that the Type~II BPS intersections will give
rise to Type~0 intersections of various $Dp_{\pm}$-branes.  It would
be nice to study the stability of these Type~0 intersections.
%%%%%%%%%%%%%%%%%%%%%%%%%%%%%%%%%%%%%%%%%%%%%%%%%%%%%%%%%%%%%%%%%%%%%%
%%%%%%%%%%%%%%%%%%%%%%%%%%%%%%%%%%%%%%%%%%%%%%%%%%%%%%%%%%%%%%%%%%%%%%
%%%%%%%%%%%%%%%%%%%%%%%%%%%%%%%%%%%%%%%%%%%%%%%%%%%%%%%%%%%%%%%%%%%%%%
%%%%%%%%%%%%%%%%%%%%%%%%%%%%%%%%%%%%%%%%%%%%%%%%%%%%%%%%%%%%%%%%%%%%%%
\section{Conclusion}
In this paper we have, starting from the Type~0B effective action,
constructed the Type~0A effective action by means of T-duality.
Although there is a doubling of RR-fields in the Type~0 theories
w.r.t.  the Type~II theories, T-duality does not mix RR-fields from
the different sectors.  Due to this doubling of RR-fields, one can
write down a democratic formulation of the action, in which we dualize
the fields of one sector giving rise to an action with only kinetic
terms for the RR-fields, {\em i.e.} in the democratic formulation
there is no Chern-Simons term in the action. Using this democratic
formulation we applied generalized dimensional reduction based on the
translational symmetry of the RR-scalar(s), in order to find the
Type~0 analogue of Romans' theory, massive Type~0A.
\par
Type~0 inherits a $\mathbb{Z}_{2}$ symmetry from the left worldsheet
fermion number operator, which takes the tachyon to its negative and
interchanges the electric ($^{+}$) and magnetic ($^{-})$ RR-sector.
This discrete symmetry together with the consistency of T-duality
between the Type~0 effective actions, then constrains the possible
couplings of the tachyon to the RR-fields.
\par
Finally we have shown how to create Type~0 solutions starting from
Type~II solutions, assuming a constant tachyon. In short, it all boils
down to change a Type~II $Dp$-brane to a Type~0 $Dp_{\pm}$- brane,
which is nothing but a bound state of a $Dp_{+}$- and a
$Dp_{-}$-brane. In particular we can embed the Type~II BPS
intersections.  For these intersections the stability is guaranteed by
supersymmetry which to a large extent also dictates the form of the
solution.  Now the form of the Type~0 intersections is the same,
making the question of stability in this class of solution an
interesting one.
%%%%%%%%%%%%%%%%%%%%%%%%%%%%%%%%%%%%%%%%%%%%%%%%%%%%%%%%%%%%%%%%%%%%%%
%%%%%%%%%%%%%%%%%%%%%%%%%%%%%%%%%%%%%%%%%%%%%%%%%%%%%%%%%%%%%%%%%%%%%%
%%%%%%%%%%%%%%%%%%%%%%%%%%%%%%%%%%%%%%%%%%%%%%%%%%%%%%%%%%%%%%%%%%%%%%
%%%%%%%%%%%%%%%%%%%%%%%%%%%%%%%%%%%%%%%%%%%%%%%%%%%%%%%%%%%%%%%%%%%%%%
\section*{Acknowledgments}
The authors would like to thank Frederik Roose for a great many
discussions.
The work of P.M.~was partially supported by the F.W.O.-Vlaanderen and
the E.U.~RTN program HPRN-CT-2000-00131.  The work of T.O.~has been
supported in part by the Spanish grant FPA2000-1584.
%%%%%%%%%%%%%%%%%%%%%%%%%%%%%%%%%%%%%%%%%%%%%%%%%%%%%%%%%%%%%%%%%%%%%%
%%%%%%%%%%%%%%%%%%%%%%%%%%%%%%%%%%%%%%%%%%%%%%%%%%%%%%%%%%%%%%%%%%%%%%
%%%%%%%%%%%%%%%%%%%%%%%%%%%%%%%%%%%%%%%%%%%%%%%%%%%%%%%%%%%%%%%%%%%%%%
%%%%%%%%%%%%%%%%%%%%%%%%%%%%%%%%%%%%%%%%%%%%%%%%%%%%%%%%%%%%%%%%%%%%%%
%\appendix
%%%%%%%%%%%%%%%%%%%%%%%%%%%%%%%%%%%%%%%%%%%%%%%%%%%%%%%%%%%%%%%%%%%%%%
%%%%%%%%%%%%%%%%%%%%%%%%%%%%%%%%%%%%%%%%%%%%%%%%%%%%%%%%%%%%%%%%%%%%%%
%%%%%%%%%%%%%%%%%%%%%%%%%%%%%%%%%%%%%%%%%%%%%%%%%%%%%%%%%%%%%%%%%%%%%%
%%%%%%%%%%%%%%%%%%%%%%%%%%%%%%%%%%%%%%%%%%%%%%%%%%%%%%%%%%%%%%%%%%%%%%

%%%%%%%%%%%%%%%%%%%%%%%%%%%%%%%%%%%%%%%%%%%%%%%%%%%%%%%%%%%%%%%%%%%%%%%%%%%%%
%
%   That's all Folks!
%
%%%%%%%%%%%%%%%%%%%%%%%%%%%%%%%%%%%%%%%%%%%%%%%%%%%%%%%%%%%%%%%%%%%%%%%%%%%%%
\end{document}